\documentclass[pra,showpacs,floatfix]{revtex4}
\usepackage{amssymb}
\usepackage{graphicx}

\newcommand{\beq}{\begin{equation}}
\newcommand{\eeq}{\end{equation}}
\newcommand{\beqa}{\begin{eqnarray}}
\newcommand{\eeqa}{\end{eqnarray}}
\newcommand{\ba}{\begin{array}}
\newcommand{\ea}{\end{array}}

\begin{document}

\title{Two-dimensional discrete solitons in dipolar Bose-Einstein condensates%
}
\author{Goran Gligori\'c$^1$, Aleksandra Maluckov$^2$, Milutin Stepi\'c$^1$,
Ljup\v co Had\v zievski$^1$, and Boris A. Malomed$^3$}
\affiliation{$^1$ Vin\v ca Institute of Nuclear Sciences, P.O. Box 522,11001 Belgrade,
Serbia \\
$^2$ Faculty of Sciences and Mathematics, University of Ni\v s, P.O. Box
224, 18001 Ni\v s, Serbia \\
$^3$ Department of Physical Electronics, School of Electrical Engineering,
Faculty of Engineering, Tel Aviv University, Tel Aviv 69978, Israel}

\begin{abstract}
We analyze the formation and dynamics of bright unstaggered solitons in the
disk-shaped dipolar Bose-Einstein condensate, which features the interplay
of contact (collisional) and long-range dipole-dipole (DD) interactions
between atoms. The condensate is assumed to be trapped in a strong
optical-lattice potential in the disk's plane, hence it may be approximated
by a two-dimensional (2D) discrete model, which includes the on-site
nonlinearity and cubic long-range (DD) interactions between sites of the
lattice. We consider two such models, that differ by the form of the on-site
nonlinearity, represented by the usual cubic term, or more accurate
nonpolynomial one, derived from the underlying 3D Gross-Pitaevskii equation.
Similar results are obtained for both models. The analysis is focused on
effects of the DD interaction on fundamental localized modes in the lattice
(2D discrete solitons). The repulsive isotropic DD nonlinearity extends the
existence and stability regions of the fundamental solitons. New families of
on-site, inter-site and hybrid solitons, built on top of a finite
background, are found as a result of the interplay of the isotropic
repulsive DD interaction and attractive contact nonlinearity. By themselves,
these solutions are unstable, but they evolve into robust breathers which
exist on an oscillating background. In the presence of the repulsive contact
interactions, fundamental localized modes exist if the DD interaction
(attractive isotropic or anisotropic) is strong enough. They are stable in
narrow regions close to the anticontinuum limit, while unstable solitons
evolve into breathers. In the latter case, the presence of the background is
immaterial.
\end{abstract}

\maketitle

\section{Introduction}

Disk-shaped two-dimensional (2D) Bose-Einstein condensates (BECs) have been
created using a superposition of a tight optical trap, formed by a pair of
parallel strongly repelling (blue-detuned) light sheets, and a loose
in-plane radial magnetic trap \cite{refer1}. Such configurations may exist
with both self-repulsive and self-attractive collisional nonlinearity, which
can be controlled by means of the Feshbach-resonance technique. In the
low-density limit, the dynamics of the 2D condensate is modeled by the 2D
Gross-Pitaevskii equation (GPE) with the cubic nonlinearity. If the density
is not very small, a consistent reduction of the underlying
three-dimensional cubic GPE to two dimensions leads to the 2D nonpolynomial
nonlinear Schr\"{o}dinger equation (NPSE) \cite{luca,luca2dnovi} (a related
but different approach leads to another form of the NPSEs for 1D and 2D
configurations, in the case of the self-repulsive nonlinearity \cite{Canary}%
).

A strong 2D periodic optical-lattice (OL) potential, applied to the
disk-shaped BEC, splits it into a planar array of droplets coupled by weak
nearest-neighbor linear interactions, due to the tunneling of atoms across
potential barriers which separate the droplets \cite{ChaosReview}. In this
situation, the BEC can be modeled by the 2D discrete GPE, derived from its
continuous counterpart by means of an expansion based on localized Wannier
functions \cite{wannier,pgk2dfund}. The same discretization procedure, if
applied to the continuous equation with the nonpolynomial nonlinearity,
leads to the discrete variant of the NPSE, as recently demonstrated in
detail in the 1D geometry \cite{Luca+we}. A noteworthy feature of the NPSE
models, both continuous and discrete ones, is that, in the case of the
attractive local nonlinearity, they account for the onset of the collapse
even in the framework of the 1D description (recall that the GPE with the
self-focusing cubic term does not give rise to the collapse in 1D).

The BEC dynamics is strongly affected by long-range interactions in \textit{%
dipolar condensates}, which can be formed by chromium atoms in an external
magnetic field, as shown in experiments \cite{Cr}. In particular, the
dipole-dipole (DD) attraction in the condensate gives rise to a specific ($d$%
-wave) mode of the collapse in the condensate \cite{d-collapse}.
Nevertheless, the dipolar condensate can be stabilized against the collapse,
adjusting the scattering length of the collisional interactions via the
Feshbach-resonance technique \cite{experim1}. It was also proposed to create
a condensate dominated by DD interactions between electric dipole moments,
that may be induced in atoms by an external electric field \cite{dc}. A
similar situation may be expected in BEC formed by dipolar molecules \cite%
{hetmol}. In particular, the creation of LiCs dipolar molecules in an
ultracold gas mixture has been reported \cite{LiCs}.

Once dipolar condensates are available to the experiment, a natural issue is
how the nonlocal DD interactions affects localized modes (solitons). Various
possibilities to create 2D solitons in dipolar condensates were explored
theoretically. It was demonstrated that isotropic solitons \cite{Pedri05}
and vortices \cite{Ami2} may exist in the 2D disk-shaped configuration with
the moments polarized perpendicular to the disk's plane, if the sign of the
DD interaction is inverted by means of the rapid rotation of dipoles \cite%
{reversal}. The ordinary DD interaction (with the uninverted sign) can give
rise to stable anisotropic 2D solitons, if the dipoles are polarized in the
plane of the confining disk \cite{Ami1}. It may be relevant to mention that
stable 2D solitons supported by nonlocal interactions (different form those
corresponding to the DD forces) are known in optical media with the thermal
nonlinearity \cite{Krolik}.

Solitons in 1D models of the dipolar BEC have been recently predicted too,
using the respective GPE \cite{Cuevas} or NPSE \cite{we-latest}. In the
latter case, the model enables to study the influence of the long-range DD
nonlinearity on the onset of the collapse in the condensate with the
self-attractive nonlinearity. Stable 1D solitons were also predicted in the
model of the Tonks-Girardeau (TG)\ gas of bosons carrying permanent moments,
using the known model of the TG gas with the self-repulsive quintic local
term, to which the long-range DD attraction was added \cite{BBB}.

A natural extension of the theoretical studies of localized modes in the
dipolar BEC is to include a strong OL potential, which leads to the discrete
model with the DD interactions between remote lattice sites. Recently, this
analysis was performed in the 1D setting \cite{gpe,Santos,NPSE}. In
particular, unstaggered solitons in such models, with both the cubic and
nonpolynomial on-site nonlinearities, were studied, respectively, in Refs.
\cite{gpe} and \cite{NPSE}, with a conclusion that the DD interactions might
enhance the stability of the discrete solitons.

The objective of the present work is to introduce discrete models for the
dipolar condensate trapped in a deep 2D lattice, and analyze the existence
and stability of localized modes in these settings (discrete bright
solitons). We will focus on the most fundamental case of \textit{unstaggered}
solitons, performing the analysis in terms of two different lattice models.
The first model includes the ordinary cubic on-site nonlinearity, i.e., it
corresponds to the 2D version of the cubic discrete GPE (to be abbreviated
as DGPE), while the second model deals with the nonpolynomial local
nonlinearity, which corresponds to the 2D discrete NPSE (alias DNPSE). Both
models incorporate the cubic nonlocal term accounting for the DD
interactions. The results obtained for discrete solitons in the frameworks
of both models will be compared.

The paper is structured as follows. The two discrete models are formulated
in Section II. Families of on-site, inter-site and hybrid 2D solitons in the
case of the attractive contact interaction, combined with isotropic and
anisotropic DD interactions, are presented in Section III. Their stability
is analyzed in the same section. Properties of fundamental localized
structures that may be formed as a result of the competition between the
repulsive local nonlinearity and attractive DD isotropic interactions (i.e.,
with the inverted sign of the DD force, as mentioned above), as well as
between the repulsive local and anisotropic DD interactions with the natural
sign, are considered in Section IV. The paper is concluded by Section V.

\section{The models}

The scaled form of the continuous GPE, which includes both contact and DD
interactions in the 3D geometry, is well known \cite{Cr,Pedri05}:
\begin{equation}
i\frac{\partial \psi \left( \mathbf{r}\right) }{\partial t}=\left[ -\frac{1}{%
2}\nabla _{\mathbf{r}}^{2}+2\pi \gamma \left\vert \psi (\mathbf{r}%
)\right\vert ^{2}+g\int \frac{1-3\cos ^{2}\theta }{\left\vert \mathbf{r-r}%
^{\prime }\right\vert ^{3}}\left\vert \psi (\mathbf{r}^{\prime })\right\vert
^{2}d\mathbf{r}^{\prime }+U(\mathbf{r})\right] \psi \left( \mathbf{r}\right)
,  \label{GPE}
\end{equation}%
where $U(\mathbf{r})$ is the external potential, $\gamma \equiv
2Na_{s}/a_{z} $ is the normalized strength of the local interactions, $N$
the number of bosonic atoms, $a_{s}$ the scattering length of inter-atomic
collisions, and $a_{z}=\sqrt{\hbar /\left( m\omega _{z}\right) }$ is the
scale of the trapping provided by potential $\left( 1/2\right) m\omega
^{2}z^{2}$, which acts perpendicular to plane $\left( x,y\right) $ of the
disk-shaped condensate. The total potential, $U(\mathbf{r})$, includes both
the latter term and the in-plane 2D OL potential, $W(x,y)$. The notation
used in Eq. (\ref{GPE}) assumes that wave function $\psi \left( \mathbf{r}%
,t\right) $ is subject to the usual normalization,
\begin{equation}
\int \left\vert \psi \left( \mathbf{r},t\right) \right\vert ^{2}d\mathbf{r}%
=1.  \label{normal}
\end{equation}%
Coefficient $g$ in Eq. (\ref{GPE}) defines the strength and sign of the DD
interactions between atomic dipoles, and $\theta $ is the angle between
vector $\left( \mathbf{r}-\mathbf{r}^{\prime }\right) $ and the orientation
of dipoles, which is fixed by a strong external magnetic field.

The reduction of Eq. (\ref{GPE}) to the 2D form is performed by assuming \
the factorization of the 3D wave function \cite{luca,luca2dnovi}:
\begin{equation}
\psi \left( \mathbf{r},t\right) =\frac{1}{\sqrt{2\pi \left\vert \gamma
\right\vert }}\phi \left( x,y,t\right) f\left( z,\eta \right) ,  \label{fac}
\end{equation}%
where $\eta $ is the width of the distribution along axis $z$ ($\eta $ may
depend on coordinates $x,y$ and time $t$), $\phi \left( x,y,t\right) $ is an
arbitrary in-plane (2D) wave function, and $f\left( z;\eta \right) $ is its
normalized axial counterpart,
\begin{equation}
f\left( z;\eta \right) =\left( \pi \eta ^{2}\right) ^{-1/4}\exp \left[
-z^{2}/\left( 2\eta ^{2}\right) \right] ,~\int_{-\infty }^{+\infty
}f^{2}\left( z\right) dz=1.  \label{norfac}
\end{equation}%
With regard to the underlying 3D normalization condition (\ref{normal}), the
factorized ansatz based on Eqs. (\ref{fac}) and (\ref{norfac}) yields the
following normalization condition for the 2D wave function,
\begin{equation}
\int \int \left\vert \phi \left( x,y\right) \right\vert ^{2}dxdy=\sqrt{2\pi
\left\vert \gamma \right\vert }.  \label{normal2}
\end{equation}

Substituting the factorized expression (\ref{fac}) in Eq. (\ref{GPE}),
assuming the presence of the strong 2D optical lattice potential in the
plane of $\left( x,y\right) $, and averaging the result along coordinate $z$%
, the following 2D NPSE with the nonlocal term accounting for the DD
interactions is derived:
\begin{eqnarray}
i\frac{\partial \phi }{\partial t} &=&\left[ -\frac{1}{2}\nabla _{\bot
}^{2}+W\left( x,y\right) +\frac{\aleph \left\vert \phi \right\vert ^{2}}{%
\eta }+\frac{1}{2}\left( \frac{1}{\eta ^{2}}+\eta ^{2}\right) \right.
\nonumber \\
&&\left. +\Gamma \int \int \frac{1-3\cos ^{2}\theta }{\left\vert \left(
x-x^{\prime }\right) ^{2}+\left( y-y^{\prime }\right) ^{2}\right\vert ^{3/2}}%
\left\vert \phi \left( x^{\prime },y^{\prime }\right) \right\vert
^{2}dx^{\prime }dy^{\prime }\right] \phi ,  \label{TDNPSE}
\end{eqnarray}%
where $\aleph $ is the sign of the contact interaction ($\aleph =-1$ and $+1$
correspond for the self-attraction and repulsion, respectively), and $\Gamma
\equiv g/\left( \sqrt{2\pi }\left\vert \gamma \right\vert \right) $ is the
relative strength of the nonlocal (DD) interaction versus its local
counterpart which can take positive and negative values. In particular, the
nonlocal term in Eq. (\ref{TDNPSE}) is obtained by the integration in the
direction of $z$, assuming the long-distance limit \cite{Fisher, Parker},
i.e., that characteristic length scales in the $\left( x,y\right) $ plane
are essentially larger than the above-mentioned transverse-confinement size,
$a_{z}$.

In the presence of the strong in-plane OL potential, that will make it
possible to reduce the model to the discrete form (see below), the latter
condition amounts to the assumption that $a_{z}$ is essentially smaller that
the lattice period. For instance, the trapping frequency $\omega _{z}=5$ kHz
yields $a_{z}\simeq 0.5$ $\mathrm{\mu }$m for chromium atoms, while the
OL period can be made equal to $5$ $\mathrm{\mu }$m, hence the latter
condition can be readily satisfied. Actually, this means that the DD
coupling may be considered as interaction between dipolar droplets with
collective magnetic moments, trapped in local potential wells of the OL \cite%
{gpe,Santos}, while inside the strongly confined droplets the DD interaction
effectively reduces to a contact form \cite{Sanlew}

Continuing the derivation of the 2D model, axial width $\eta $, introduced
above in Eq. (\ref{norfac}), is determined, in the lowest approximation by a
local relation which reduces to an algebraic equation,
\begin{equation}
\eta ^{4}=1+\aleph \left\vert \phi \right\vert ^{2}\eta  \label{alg}
\end{equation}%
\cite{luca}.

In the low-density limit, $\left\vert \phi \right\vert ^{2}<<1$, Eq. (\ref%
{TDNPSE}) is tantamount to the GPE with the ordinary cubic local
nonlinearity:
\begin{eqnarray}
i\frac{\partial \phi }{\partial t} &=&\left[ -\frac{1}{2}\nabla _{\bot
}^{2}+W\left( x,y\right) +\aleph \left\vert \phi \right\vert ^{2}\right.
\nonumber \\
&&\left. +\Gamma \int \int \frac{1-3\cos ^{2}\theta }{\left\vert \left(
x-x^{\prime }\right) ^{2}+\left( y-y^{\prime }\right) ^{2}\right\vert ^{3/2}}%
\left\vert \phi \left( x^{\prime },y^{\prime }\right) \right\vert
^{2}dx^{\prime }dy^{\prime }\right] \phi .  \label{TDGPE}
\end{eqnarray}%
In this limit, Eq. (\ref{alg}) amounts to $\eta \approx 1$.

The choice of potential $W(x,y)$ in the form corresponding to a deep OL
suggests to approximate the wave function by a superposition of localized
Wannier modes, $\Phi _{m,n}\left( x,y\right) $ \cite{wannier}, $\phi
(x,y)=\sum_{m,n}\phi _{m,n}\Phi _{m,n}\left( x,y\right) $. Projecting, as
usual \cite{Luca+we}, equation (\ref{TDNPSE}) onto the set of this modes,
Eq. (\ref{TDNPSE}) is reduced to the set of equations
\begin{eqnarray}
i\frac{\partial \phi _{m,n}}{\partial t} &=&-\frac{C}{2}\left( \phi
_{m+1,n}+\phi _{m-1,n}+\phi _{m,n+1}+\phi _{m,n-1}-4\phi _{m,n}\right) +%
\left[ \frac{\aleph \left\vert \phi _{m,n}\right\vert ^{2}}{\eta _{m,n}}+%
\frac{1}{2}\left( \frac{1}{\eta _{m,n}^{2}}+\eta _{m,n}^{2}\right) \right.
\nonumber \\
&&\left. +\Gamma \sum_{m^{\prime },n^{\prime }}\frac{1-3\cos ^{2}\theta
_{m^{\prime },n^{\prime }}}{\left\vert \left( m-m^{\prime }\right)
^{2}+\left( n-n^{\prime }\right) ^{2}\right\vert ^{3/2}}\left\vert \phi
_{m^{\prime },n^{\prime }}\right\vert ^{2}\right] \phi _{m,n},  \label{DNPSE}
\end{eqnarray}%
where $\left( m,n\right) $ play the role of 2D discrete coordinates
replacing $\left( x,y\right) $, the summation in the DD part excludes the
term with $\left( m^{\prime },n^{\prime }\right) =\left( m,n\right) $, and $C
$ is an effective lattice coupling constant, determined by the overlap
integral of functions $\Phi _{m.n}\left( x,y\right) $ centered at two
adjacent sites of the lattice. In addition to Eq. (\ref{DNPSE}), discrete
width function $\eta _{m,n}$ obeys the respective counterpart of algebraic
equation (\ref{alg}),
\begin{equation}
\eta _{m,n}^{4}=1+\aleph \left\vert \phi _{m,n}\right\vert ^{2}\eta _{m,n}.
\label{dalg}
\end{equation}%
On the other hand, the 2D DGPE, which represents the discretization of Eq. (%
\ref{TDGPE}), amounts to a single equation \cite{pgk2dfund},
\begin{eqnarray}
i\frac{\partial \phi _{m,n}}{\partial t} &=&-\frac{C}{2}\left( \phi
_{m+1,n}+\phi _{m-1,n}+\phi _{m,n+1}+\phi _{m,n-1}-4\phi _{m,n}\right) +%
\left[ \aleph \left\vert \phi _{m,n}\right\vert ^{2}\right.   \nonumber \\
&&\left. +\Gamma \sum_{m^{\prime },n^{\prime }}\frac{1-3\cos ^{2}\theta
_{m^{\prime },n^{\prime }}}{\left\vert \left( m-m^{\prime }\right)
^{2}+\left( n-n^{\prime }\right) ^{2}\right\vert ^{3/2}}\left\vert \phi
_{m^{\prime },n^{\prime }}\right\vert ^{2}\right] \phi _{m,n}.  \label{DGPE}
\end{eqnarray}

Stationary solutions to Eqs. (\ref{DNPSE}) and (\ref{DGPE}) with chemical
potential $\mu $ are sought for as $\phi _{m,n}=u_{m,n}\exp \left( -i\mu
t\right) $, with discrete function $u_{m,n}$ satisfying, respectively, the
following stationary equations:
\begin{eqnarray}
\mu u_{m,n} &=&-\frac{C}{2}\left(
u_{m+1,n}+u_{m-1,n}+u_{m,n+1}+u_{m,n-1}-4u_{m,n}\right) +\left[ \frac{\aleph
\left\vert u_{m,n}\right\vert ^{2}}{\eta _{m,n}}+\frac{1}{2}\left( \frac{1}{%
\eta _{m,n}^{2}}+\eta _{m,n}^{2}\right) \right.  \nonumber \\
&&\left. +\Gamma \sum_{m^{\prime },n^{\prime }}\frac{1-3\cos ^{2}\theta
_{m^{\prime },n^{\prime }}}{\left\vert \left( m-m^{\prime }\right)
^{2}+\left( n-n^{\prime }\right) ^{2}\right\vert ^{3/2}}\left\vert
u_{m^{\prime },n^{\prime }}\right\vert ^{2}\right] u_{m,n}~;  \label{TdSNPSE}
\end{eqnarray}%
\begin{eqnarray}
\mu u_{m,n} &=&-\frac{C}{2}\left(
u_{m+1,n}+u_{m-1,n}+u_{m,n+1}+u_{m,n-1}-4u_{m,n}\right) +\left[ \aleph
\left\vert u_{m,n}\right\vert ^{2}\right.  \nonumber \\
&&\left. +\Gamma \sum_{m^{\prime },n^{\prime }}\frac{1-3\cos ^{2}\theta
_{m^{\prime },n^{\prime }}}{\left\vert \left( m-m^{\prime }\right)
^{2}+\left( n-n^{\prime }\right) ^{2}\right\vert ^{3/2}}\left\vert
u_{m^{\prime },n^{\prime }}\right\vert ^{2}\right] u_{m,n}~.  \label{TdSGPE}
\end{eqnarray}%
In the former case, $\eta _{m,n}$ are related to $u_{m,n}$ by Eq. (\ref{dalg}%
), with $\left\vert \phi _{m,n}\right\vert ^{2}$ replaced by $\left\vert
u_{m,n}\right\vert ^{2}$.

Generally, the expression for angles $\theta _{m,n}$ in the term which
describes the long-range DD interaction in the above discrete equations is
cumbersome. However, following Refs. \cite{Pedri05,Ami2,Ami1}, it makes
sense to focus on two most fundamental cases, with magnetic dipoles oriented
either parallel or perpendicular to the $\left( x,y\right) $ plane. In the
former case, we choose the orientation of the dipole moments along the $x$.
For the parallel orientation, the DD term in Eqs. (\ref{TdSNPSE}) and (\ref%
{TdSGPE}) is anisotropic, being attractive in one in-plane direction and
repulsive in the other \cite{Ami1}:
\begin{equation}
\left( \frac{1-3\cos ^{2}\theta _{m^{\prime },n^{\prime }}}{\left\vert
\left( m-m^{\prime }\right) ^{2}+\left( n-n^{\prime }\right) ^{2}\right\vert
^{3/2}}\right) _{{\LARGE \Vert }}=\frac{\left( n-n^{\prime }\right)
^{2}-2\left( m-m^{\prime }\right) ^{2}}{\left\vert \left( m-m^{\prime
}\right) ^{2}+\left( n-n^{\prime }\right) ^{2}\right\vert ^{5/2}}~.
\label{anisDD}
\end{equation}%
For the perpendicular orientation of the dipoles, the DD term is isotropic
\cite{Pedri05,Ami2}, taking a simple form, cf. the above expression:
\begin{equation}
\left( \frac{1-3\cos ^{2}\theta _{m^{\prime },n^{\prime }}}{\left\vert
\left( m-m^{\prime }\right) ^{2}+\left( n-n^{\prime }\right) ^{2}\right\vert
^{3/2}}\right) _{{\LARGE \perp }}=\frac{\left\vert \phi _{m^{\prime
},n^{\prime }}\right\vert ^{2}}{\left\vert \left( m-m^{\prime }\right)
^{2}+\left( n-n^{\prime }\right) ^{2}\right\vert ^{3/2}}~.  \label{isoDD}
\end{equation}

The character of the DD interaction in both cases is defined by the sign of $%
\Gamma $, which is related to the sign of coefficient $g$ in the underlying
GPE (\ref{GPE}). Normally, $g$ is positive. However (as mentioned above),
its sign may be reversed by means of a rapidly rotating magnetic field \cite%
{reversal}, allowing $\Gamma $ to take negative values \cite{Pedri05,Ami2}.
Thus, $\Gamma >0$ implies the repulsive isotropic DD interaction, while the
anisotropic DD interaction is repulsive along discrete coordinate $n$, and
attractive, being twice as strong in the absolute value, along $m$. For $%
\Gamma <0$, the isotropic DD interaction is attractive, while in the
anisotropic configuration the DD interaction is attractive along $n$ and
repulsive in the direction of $m$. The integral quantity that characterizes
localized discrete modes is their norm (or power, in terms of optical
models), $P=\sum_{m,n}\left\vert u_{m,n}\right\vert ^{2}$, which is the
dynamical invariant of Eqs. (\ref{DNPSE}) and (\ref{DGPE}).

For both in-plane and out-of-plane (perpendicular) dipole orientations,
stationary equations (\ref{TdSNPSE}) and (\ref{TdSGPE}) were solved by means
of a numerical algorithm based on a modified Powell minimization method,
which uses a finite-difference expression for the underlying Jacobian. In
the analysis reported below, we focus on the three most fundamental
unstaggered localized modes (\textquotedblleft staggered" are configurations
with opposite signs of the field at adjacent sites, which is relevant to
local models with the repulsive nonlinearity): \textit{on-site} modes, i.e.,
those centered on a lattice site, \textit{inter-site} ones, which are
centered between lattice sites in both directions $m$ and $n$, and \textit{%
hybrid modes}, centered on-site along one lattice direction, and between
sites along the other. This nomenclature was adopted in previous studies of
local discrete models \cite{pgk2dfund}. Results reported below were obtained
using a square lattice composed of $11$x$11$ or $21$x$21$ elements for
on-site modes, and of $10$x$10$ or $20$x$20$ elements for inter-site and
hybrid ones. It was checked that the results did not alter if an essentially
larger lattices were used.

\section{Fundamental unstaggered solitons in the case of attractive contact
interactions}

The nonlinear localization of matter waves occurs in gaps of the linear
spectrum of the underlying system. In the case of the attractive local
nonlinearity, fundamental solitons populate the semi-infinite gap \cite%
{wannier,kivshargap}, which corresponds to regions $\mu <0$ and $\mu <1$, in
the GPE and NPSE models, respectively. Discrete models, derived in the
tight-binding approximation, cannot correctly describe the entire linear
spectrum of the underlying continuum \cite{tightb}. However, they can
capture localized modes existing inside the semi-infinite gap. We will
consider fundamental localized modes in this case (their topologically
modified counterparts, such as discrete vortices \cite{discrete-vortices},
will be considered elsewhere).

To outline the entire existence region of the fundamental solitons, we
followed the known approach, identifying regions where continuous-wave (CW)
solutions with a given chemical potential, $\mu $, are subject to the
modulational instability, MI (it is assumed that the MI splits unstable CW
states into arrays of solitons) \cite{gpe}-\cite{NPSE}. Further details of
the respective numerical procedure are reported elsewhere \cite{acta}. It is
relevant to notice that the instability of uniform states in the dipolar gas
is determined by two different factors: the local self-focusing, which is
the usual driving mechanism for the MI at small wavenumbers of
perturbations, and, in addition, the roton instability at finite
wavenumbers, induced by the long-range DD interactions \cite{roton}.

\subsection{The effect of the DD interaction on fundamental solitons}

As mentioned above, in the local model with the attractive contact
interactions ($\aleph =-1$), CW solutions are modulationally unstable in
regions $\mu <1$ and $\mu <0$ for the DNPSE and DGPE models, respectively.
Note that these values correspond to the upper boundaries of the
semi-infinite gaps in the corresponding continuous NPSE and GPE models. The
presence of the nonlocal DD interaction extends the MI domain of the CW
states, thus enabling the existence of localized modes beyond the
above-mentioned limit values, $\mu =1$ and $\mu =0$.

\begin{figure}[tbp]
\center\includegraphics [width=5cm]{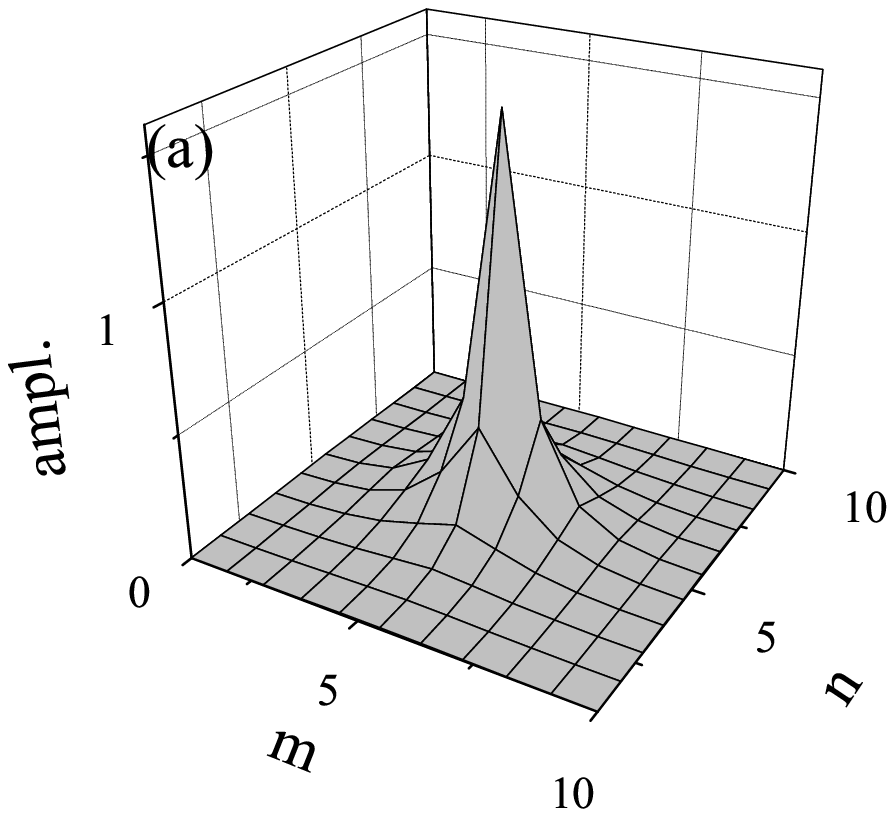}
\includegraphics [width=5cm]{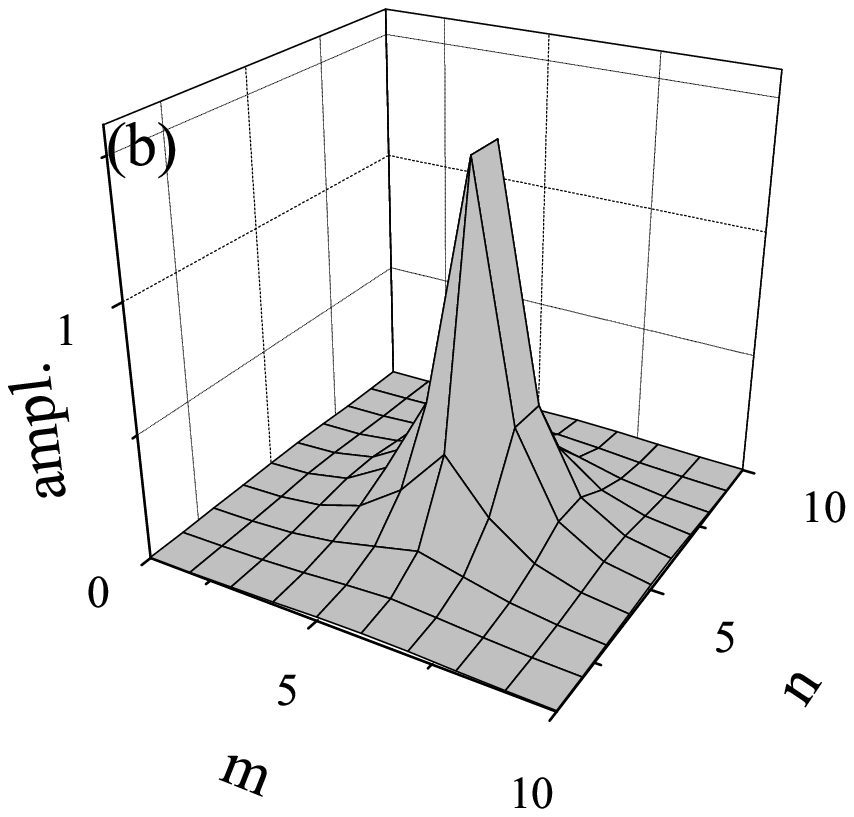}\includegraphics
[width=5cm]{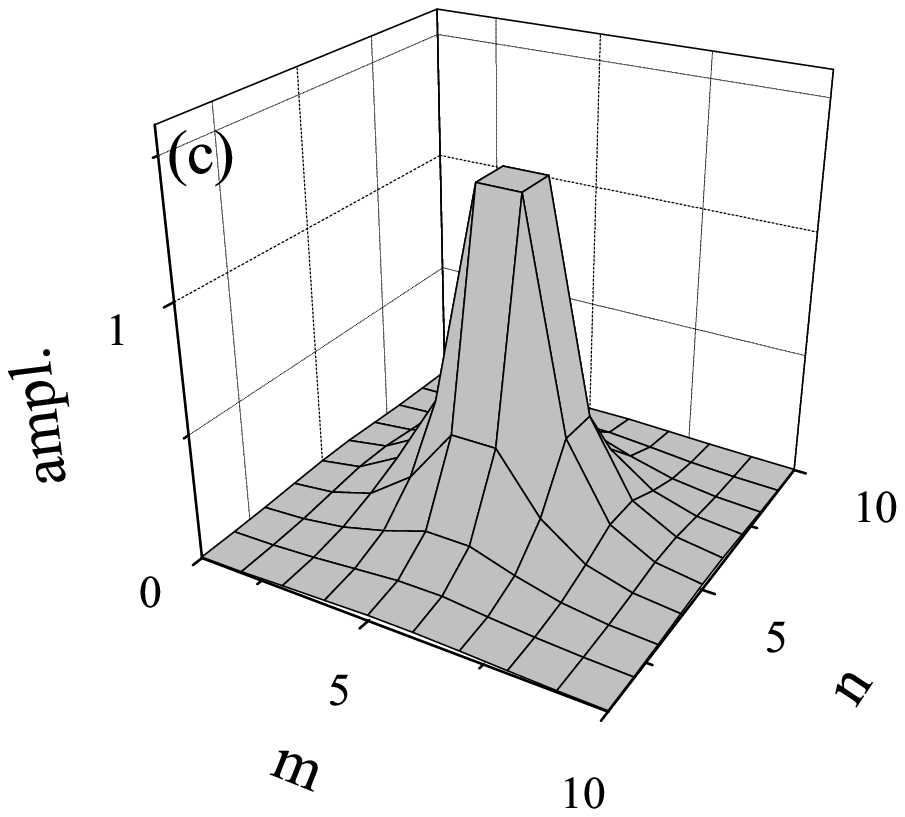} \caption{Examples of different types of
fundamental unstaggered solitons in the 2D DNPSE: on-site (a),
hybrid (b), and inter-site (c). The contact interaction is
attractive, while the DD interaction is absent, in this case. The
coupling constant is $C=2$.} \label{fig1}
\end{figure}

In the case of the attractive contact interactions, families of fundamental
unstaggered bright solitons of the three above-mentioned types -- on-site,
inter-site, and hybrid (see Fig. \ref{fig1}) -- have been found, as
expected, precisely in the domain of the parameter space featuring the MI of
the CW states. To this end, stationary equations (\ref{TdSNPSE}) and (\ref%
{TdSGPE}) were solved for the DNPSE and DGPE models, respectively.

In the case of the attractive contact interaction, the attractive
isotropic DD nonlinearity (that with the inverted sign) acts in the
same direction and cannot change qualitative properties of the
solitons. Therefore, only the case of the attractive on-site
nonlinearity competing with its repulsive DD isotropic counterpart
is really interesting, similar to the situation 1D counterpart of
the present model \cite{Cuevas}, and in the Salerno model (a
combined Ablowitz-Ladik/discrete nonlinear-Sch\"{o}dinger system)
with competition between the attractive on-site and repulsive
inter-site nonlinearities \cite{Zaragoza}. For this reason, in the
case of the attractive contact interaction, we fix $\Gamma >0$ (the
natural sign of the DD forces). In parallel, the interplay of the
attractive on-site and anisotropic DD interactions will also be
considered for $\Gamma >0$.

\begin{figure}[tbp]
\center\includegraphics [width=12.7cm]{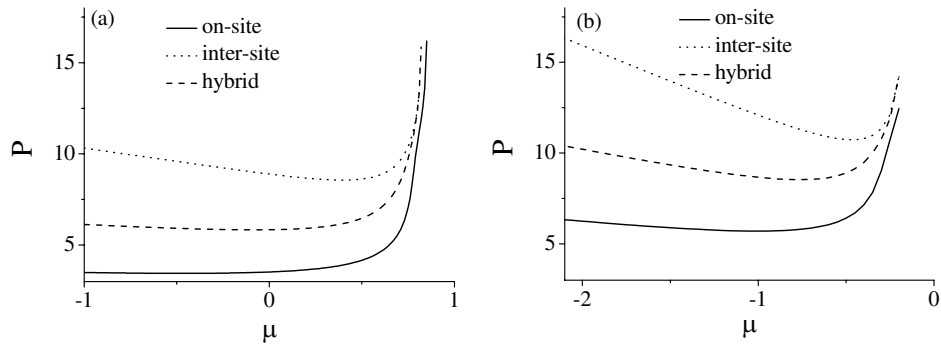} \caption{The
$P(\protect\mu )$ dependencies for on-site, hybrid and inter-site
solitons in the two-dimensional DNPSE (a) and DGPE (b) models,
without the DD interactions; $C=2$. } \label{fig2}
\end{figure}

Basic characteristics of soliton families are presented by the $P(\mu )$
dependence, which shows the norm versus the chemical potential. First, in
Fig. \ref{fig2} we display $P\left( \mu \right) $ curves generated by the
DNPSE (a) and DGPE (b) models for the on-site, hybrid and inter-site
discrete solitons in the presence of the attractive contact interaction,
while the DD interaction is still absent. These plots may be used as the
reference point for the comparison with results obtained in the presence of
the DD interactions.

\begin{figure}[tbp]
\center\includegraphics [width=7cm]{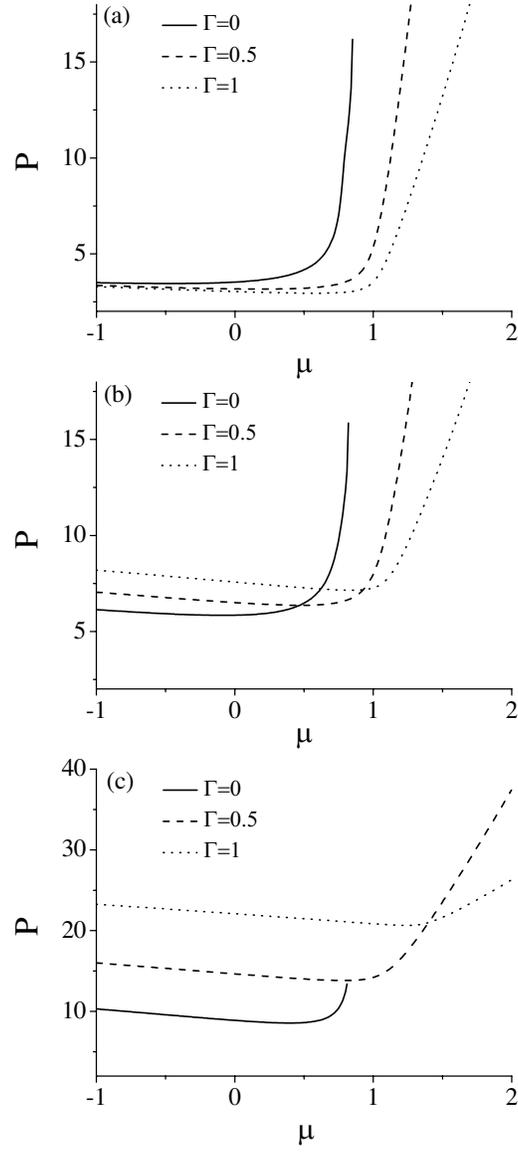} \caption{The
$P(\protect\mu )$ dependencies for families of 2D solitons of the
on-site (a), hybrid (b), and inter-site (c) types in the DNPSE
model, in
the presence of the isotropic DD interaction with relative strength $\Gamma $%
. The coupling constant is $C=2$. }
\label{fig3}
\end{figure}

\begin{figure}[tbp]
\center\includegraphics [width=7cm]{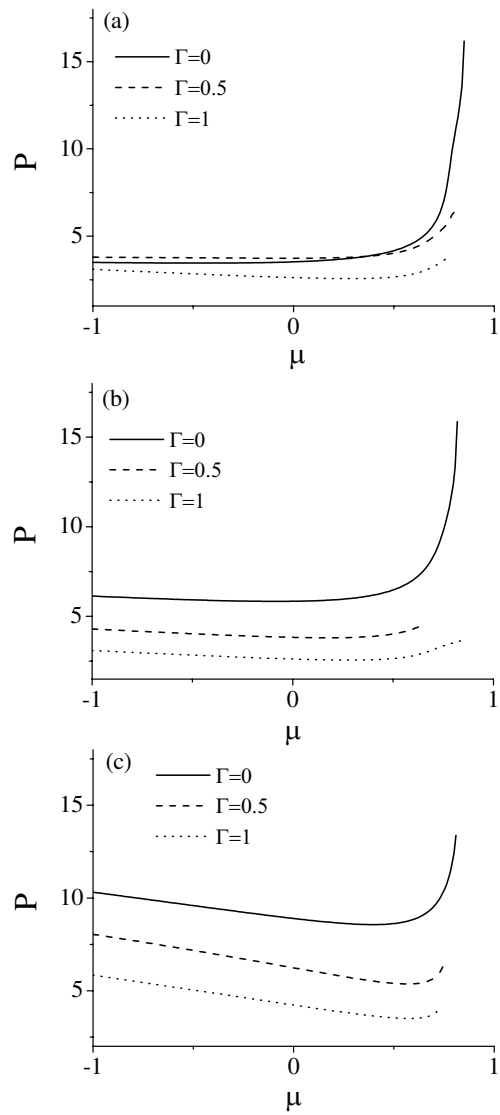}\caption{The same
as in Fig. \protect\ref{fig3}, but in the case of the anisotropic
DD interaction, which corresponds to the in-plane orientation of
dipolar moments.} \label{fig4}
\end{figure}

The $P(\mu )$ dependencies for the soliton families affected by the
isotropic (repulsive) and anisotropic DD interactions are displayed in Figs. %
\ref{fig3} and \ref{fig4}, respectively. The comparison of the $P(\mu )$
dependencies displayed in Figs. \ref{fig3} and \ref{fig4} with those in Fig. %
\ref{fig2} shows that the repulsive isotropic DD interaction slightly
extends the region in which the fundamental localized structures can be
found, which is consistent with the results of the MI analysis for the CW
solutions \cite{acta}. Simultaneously, the obtained $P\left( \mu \right) $
dependencies show that there is no qualitative difference between the DGPE
and DNPSE models in this case.

Usually, the Vakhitov-Kolokolov (VK) criterion, alias the slope condition, $%
dP/d\mu \leq 0$, is considered as a necessary condition for the stability of
soliton \cite{VK,isrl}. Although it is strictly applicable only to systems
with a local power-law nonlinearity, it may be valid in more general
situations too. Here we consider the slope criterion in both cases combining
the cubic or nonpolynomial local nonlinearity with the nonlocal
DD-interaction term. The slope criterion is violated at values of $\mu $
close to the right edge of the existence region of the fundamental solitons,
see Figs. \ref{fig2}, \ref{fig3}, and \ref{fig4}.

\begin{figure}[tbp]
\center\includegraphics [width=9cm]{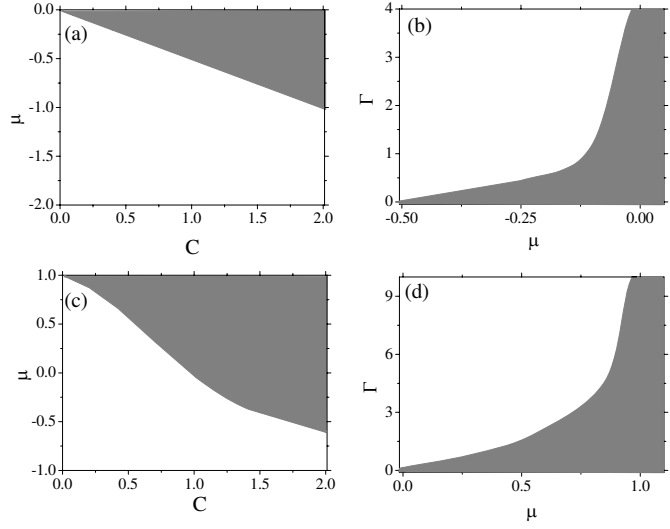}
\caption{(a,c) The stability diagram for on-site solitons in the $(\protect%
\mu ,C)$ plane in the case without the DD interaction: the DGPE (a) and NPSE
(c) two-dimensional models. The influence of the isotropic repulsive DD
interaction is presented in parts (b) and (d) for the DGPE and DNPSE models,
respectively. The lattice coupling constant is $C=1$. Solitons are stable
and unstable in white and gray areas, respectively. }
\label{fig5}
\end{figure}

The general condition which must be fulfilled for the stability of
solitons is the absence of eigenvalues with positive real parts in
the spectrum of small perturbations around the solitons (the
spectral criterion); in fact, this means the eigenvalues must have
zero real parts \cite{isrl}. This criterion was implemented
numerically by computing the corresponding eigenvalues, using
linearized equations for small perturbations, similar to how it was
done in Ref. \cite{Luca+we}. In the case of the on-site fundamental
solitons, the real part of the eigenvalues vanishes in certain
regions of parameter plane ($C$, $\mu $), as shown in Figs.
\ref{fig5} (a) and (c) for both discrete models. This indicates the
presence of a wide stability region for the on-site modes, as
reported earlier in papers dealing with 2D solitons in the discrete
nonlinear Schr\"{o}dinger equation, without DD interactions
\cite{pgk2dfund}, \cite{aniso} and \cite{rotating}. The VK criterion
is also satisfied in these cases.

In the presence of the isotropic repulsive DD interaction the region of the
existence of fundamental solitons expands. These extended regions are inside the domains where CW solutions are subject to the MI, in accordance with the above statement.
Two significant effects are generated by the nonlocal DD interaction in this
case. First, the isotropic repulsive DD interaction makes the stability
regions \emph{wider} for on-site fundamental solitons with respect to the
model without the nonlocal interaction, as seen in Figs \ref{fig5} (b) and
(d) for the DGPE and DNPSE models, respectively. Second, the nonlocal
interaction generates new localized structures rising above a \emph{finite
background}. Properties of these new stationary structures, which are
on-site centered, are considered in the following subsection. Here we only
mention that the linear stability analysis shows that these
solitons-on-the-background are unstable. On the other hand, in the presence
of the anisotropic DD interaction, the stability domain of the on-site
solitons shrinks (not shown here in detail).

In the absence of the nonlocal interactions, a known result is that the
fundamental solitons of the inter-site and hybrid types, which are
characterized by greater values of the norms in comparison with the on-site
modes, are unstable in the whole existence region \cite{pgk2dfund}. We have
checked that DD interaction of either kind (isotropic repulsive or
anisotropic) \emph{do not} stabilize these modes.

\subsection{Solitons on a finite background}

\begin{figure}[tbp]
\center\includegraphics [width=12.7cm]{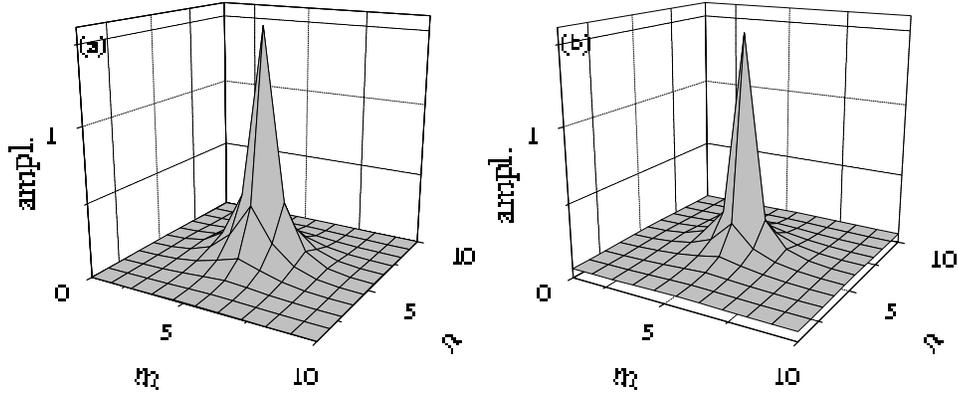}
\caption{Comparison of an ordinary (fully localized) fundamental
soliton (a) and SFB, i.e., the one sitting on a finite background
(b) in the 2D DNPSE model. The powers (norms) of both solitons are
the same, $P=3.5$. The strength of the DD interaction is $\Gamma
=0$ in case (a) and $\Gamma =1$ in case (b), while the lattice
coupling constant is the same, $C=2$. The norm of the SFB, which
includes the nonvanishing background, is finite due to the
finiteness of the lattice.} \label{fig6}
\end{figure}

\begin{figure}[tbp]
\center\includegraphics [width=9cm]{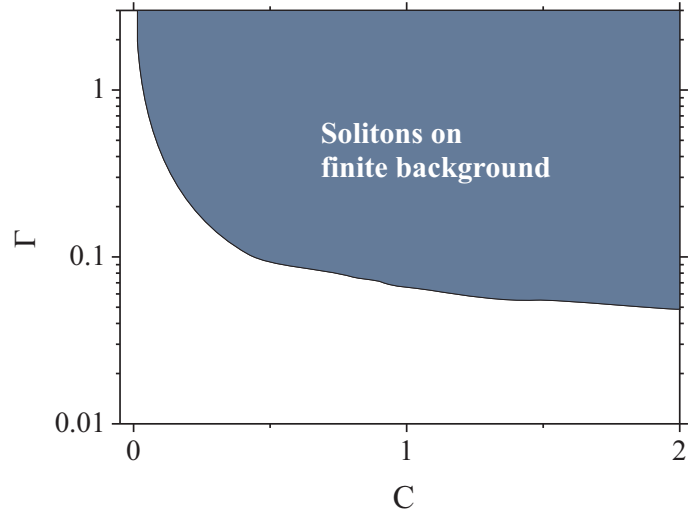} \caption{The
existence diagram of SFBs (solitons on the finite background) in
parameter plane $(C,\Gamma )$ for the combination of the
attractive contact and repulsive DD interactions in the 2D DNPSE
model. The SFB is registered if the amplitude of the background
exceeds $1/100$ of the amplitude at the center of the soliton.
With the increase of coupling
constant $C$, the SFBs appear at a smaller strength of the DD interaction, $%
\Gamma $. }
\label{fig7}
\end{figure}

As briefly mentioned above, a new feature revealed by the analysis in the
presence of the repulsive isotropic DD interactions (i.e., those with the
natural sign) is the role of the background in the formation of localized
modes. This effect is most significant near the edge of the soliton's
existence region (at $\mu $ close to $0$ or $1$ in the DGPE and DNPSE
models, respectively), where, in the presence of the isotropic DD repulsion,
narrow localized modes are found, as stationary solutions, on top of a
finite uniform background, see Fig. \ref{fig6}. We stress that such
structures cannot be found in the absence of the DD interactions. When these
interactions are present, the background may play a significant role in the
formation of the localized structures, because the background as a whole
couples to the central peak via the long-range DD forces. Actually, the long-range coupling gives rise to these newly
formed \textit{solitons on the finite background} (SFBs) when its strength
exceeds a certain threshold level, see Fig. \ref{fig7}. The threshold becomes lower with the increase of coupling constant $C$, as
seen in Fig. \ref{fig7}. Based on the waveform observed above the
background, one can identify SFBs of the on-site, inter-site, and hybrid
types.

Localized SFB\ structures are generated by the MI of a finite-amplitude CW
states in the nonlinear lattice with the DD interaction. In the course of
the development of the MI, the interplay of the local (contact) and nonlocal
(DD) nonlinearities arrests the exponential growth of the instability and
imposes global correlations, which enable the creation of the SFB as an
outcome of the nonlinear development of the MI. However, the SFBs themselves
turn out to be unstable, eventually evolving into robust breathers, which
consist of a localized vibrating peak and finite oscillating background, as
shown below.

\subsection{Dynamical properties of the fundamental localized modes}

\begin{figure}[tbp]
\center\includegraphics [width=7cm]{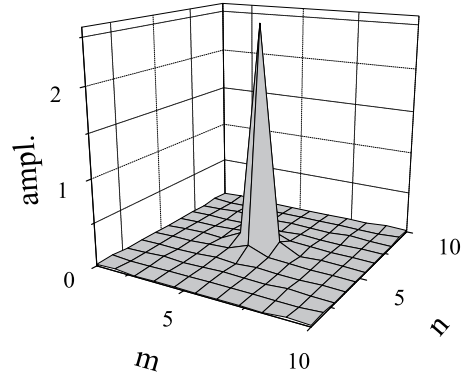} \caption{An
example of a strongly pinned on-site soliton, which remains
\textquotedblleft frozen" (completely insensitive to
perturbations) in the course of the evolution. Parameters are
$C=2,\protect\mu =-7,\Gamma =1$, which implies the combination of
the attractive contact and isotropic repulsive DD interactions. In
this and all following figures the results are presented for the
nonpolynomial model. The findings are qualitatively the same in
the model with the cubic nonlinearity.} \label{fig8}
\end{figure}

In addition to the modes described above, strongly pinned on-site
solitons are found in certain parametric areas in the model with the
attractive contact and isotropic repulsive DD interactions. Being very
narrow, these modes are not significantly affected by the DD interaction. In
general, these solutions correspond to early flat parts of the $P(\mu )$
plots in Figs. (\ref{fig2})-(\ref{fig4}) (at $\mu <-1$ and $\mu <0$ in the
DGPE and DNPSE model, respectively), and they have their counterparts ion
the local model. These localized structures, such as the one displayed in
Fig. (\ref{fig8}) for the DNPSE, remain "frozen" under the action of
arbitrary perturbations.

\begin{figure}[tbp]
\center\includegraphics [width=5cm]{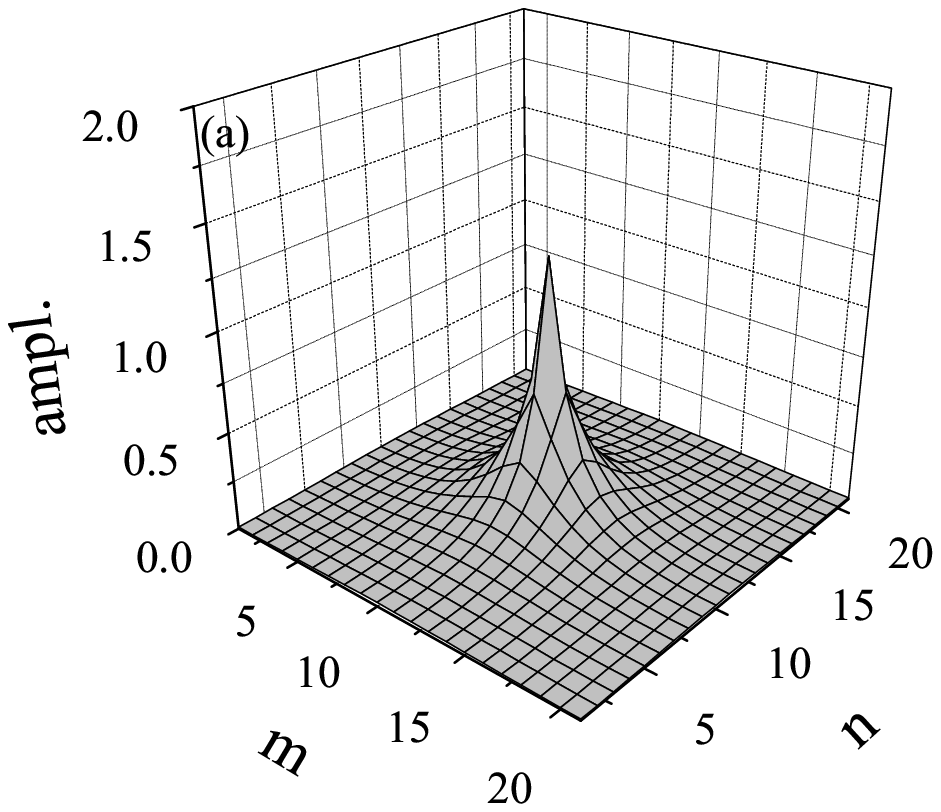}
\includegraphics [width=5cm]{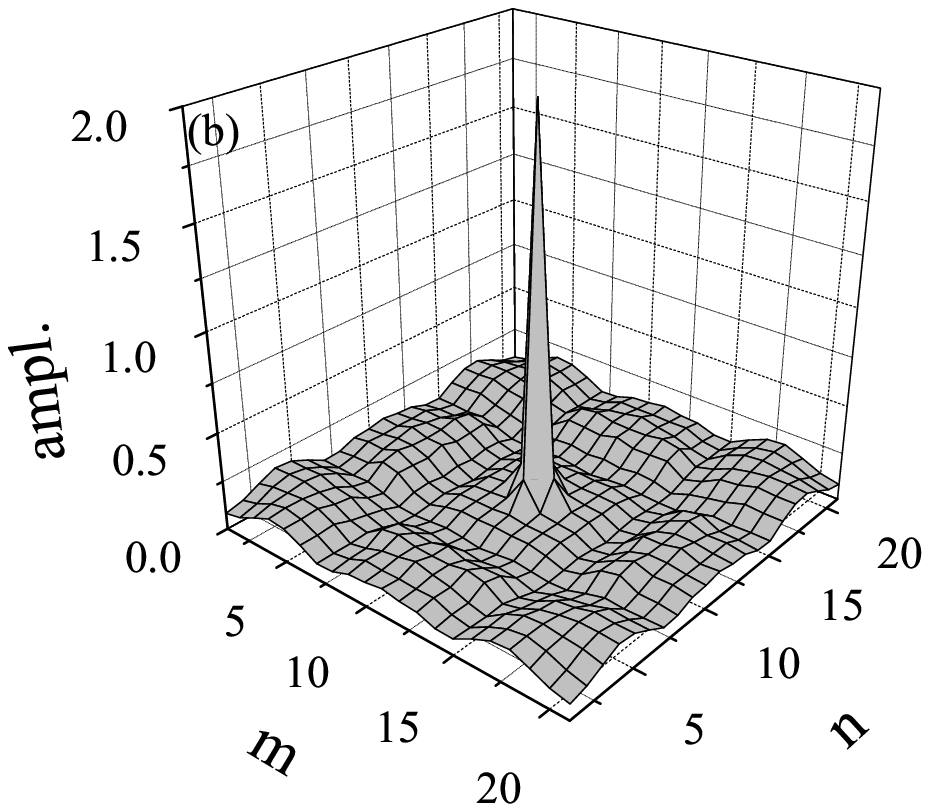}\includegraphics
[width=5cm]{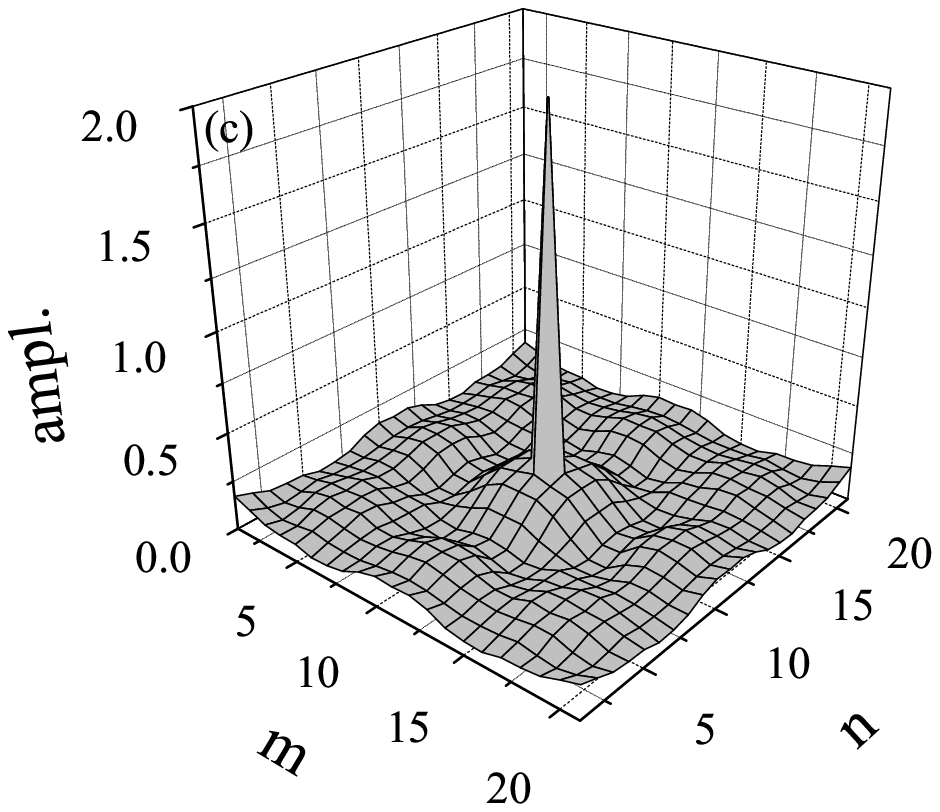}  \caption{The result of the application
of a small perturbation to an unstable on-site soliton, which
exists in the presence of the attractive contact and repulsive
isotropic DD interactions ($C=2$, $\protect\mu =0.89$, $\Gamma
=0.1$). The soliton evolves into a persistent tightly bound
structure, surrounded by an oscillating background. The snapshots
correspond to $t=1$ (a), $t=25$ (b), and $t=50$ (c).} \label{fig9}
\end{figure}

\begin{figure}[tbp]
\center\includegraphics [width=5cm]{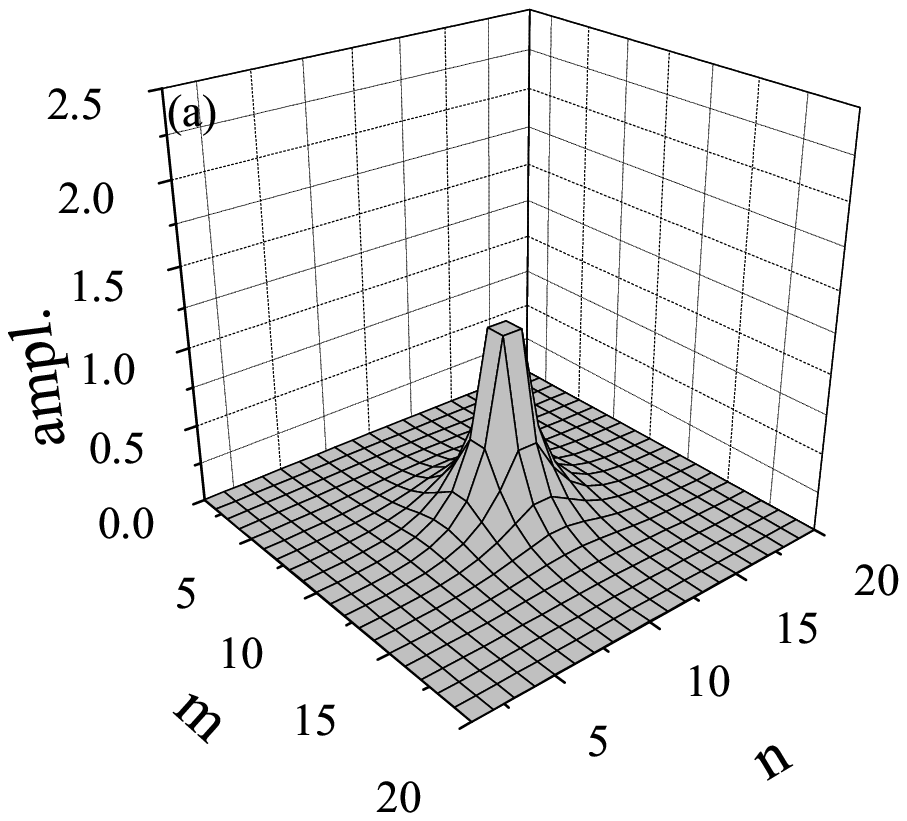}
\includegraphics [width=5cm]{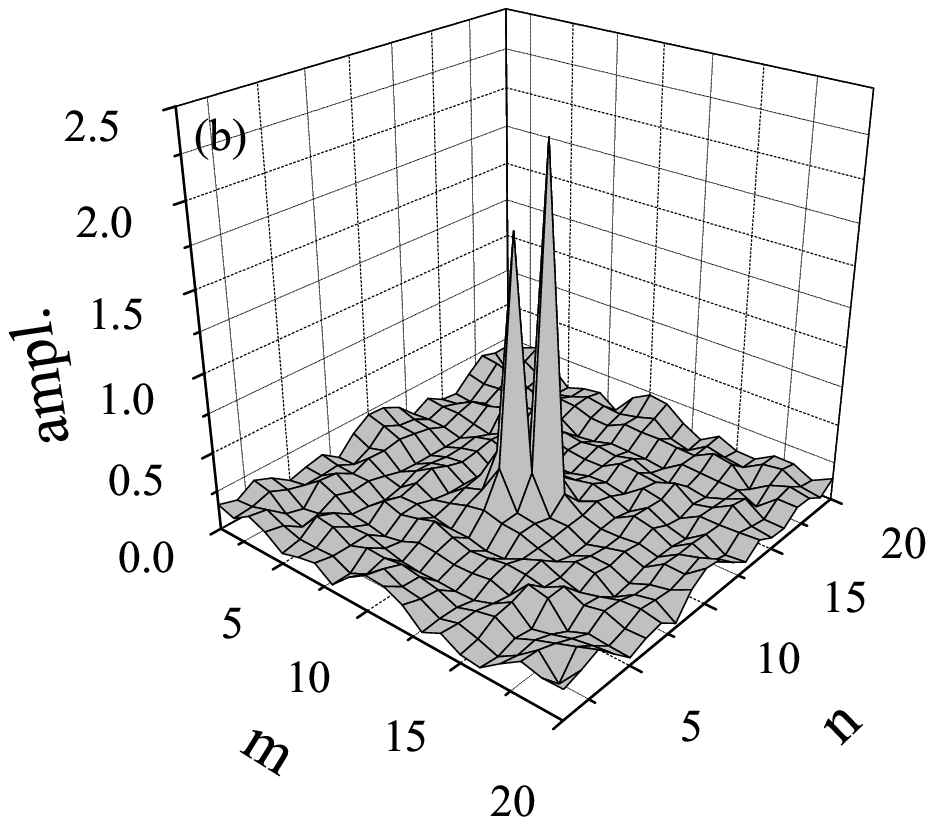}\includegraphics
[width=5cm]{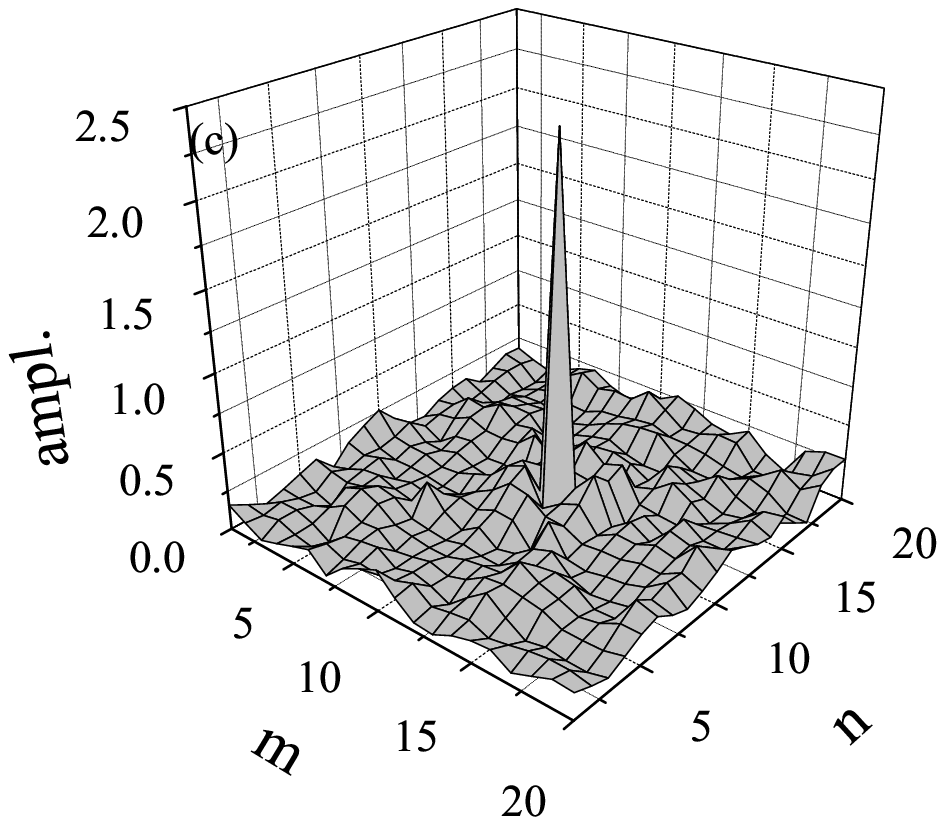}
\caption{The evolution of a perturbed inter-site soliton, for $C=0.2,\protect%
\mu =0.7,\Gamma =0.1$, i.e., the combination of the contact attraction and
isotropic DD repulsion. The emerging structure, as seen in panel (c), is a
persistent one. The snapshots pertain to $t=1$ (a), $t=25$ (b), and $t=50$
(c). }
\label{fig10}
\end{figure}

The evolution of all the unstable fundamental solitons considered in the
previous subsections follows approximately the same scenario, in the
presence of the DD interactions. Direct simulations of Eqs. (\ref{DNPSE})
and (\ref{DGPE}) demonstrate that the outcome of the evolution under the
action of perturbations is the emergence of persistent strongly pinned
on-site breathers, surrounded by a finite oscillating background, see an
example for the DNPSE in Fig. \ref{fig9}. The evolution may feature
different transient stages, depending on the initial structure. For example,
unstable on-site solitons directly evolve towards single-peaked tightly
localized breathers, while inter-site solitons temporarily form localized
structures with two peaks, see Fig. \ref{fig10}. The eventually formed
breathers are very robust and survive perturbations of any kind.

A general conclusion is that the repulsive nonlocal DD interaction in the 2D
discrete models does not prevent the formation of the extremely narrow
localized structures, which are strongly pinned to the underlying lattice.
These very narrow, dynamically stable modes have been associated with the
quasi-collapse in the 2D lattice \cite{2ddiscoll}, \cite{pitaevcollapse}.
Nevertheless, the DD interactions produce an essential effect, as they give
rise to an oscillating background which surrounds the narrow peak, being
coupled to it by the long-range DD interactions, see Figs. \ref{fig9} and %
\ref{fig10}; in the usual local model, with $\Gamma =0$, the narrow peak
would not be coupled to any background.

With the increase the strength of the DD interaction, the evolution of the
corresponding solitons of the SFB type becomes sensitive to the form of
perturbations. Under small asymmetric perturbations (which actually
correspond to small shifts of the initial soliton's location), or random
perturbations, the SFBs decay. On the other hand, under small symmetric
perturbations they evolve towards tightly bound single-peaked localized
structures, surrounded by an oscillating background, similar to those
displayed in Figs. \ref{fig9} and \ref{fig10}.

\begin{figure}[tbp]
\center\includegraphics [width=5cm]{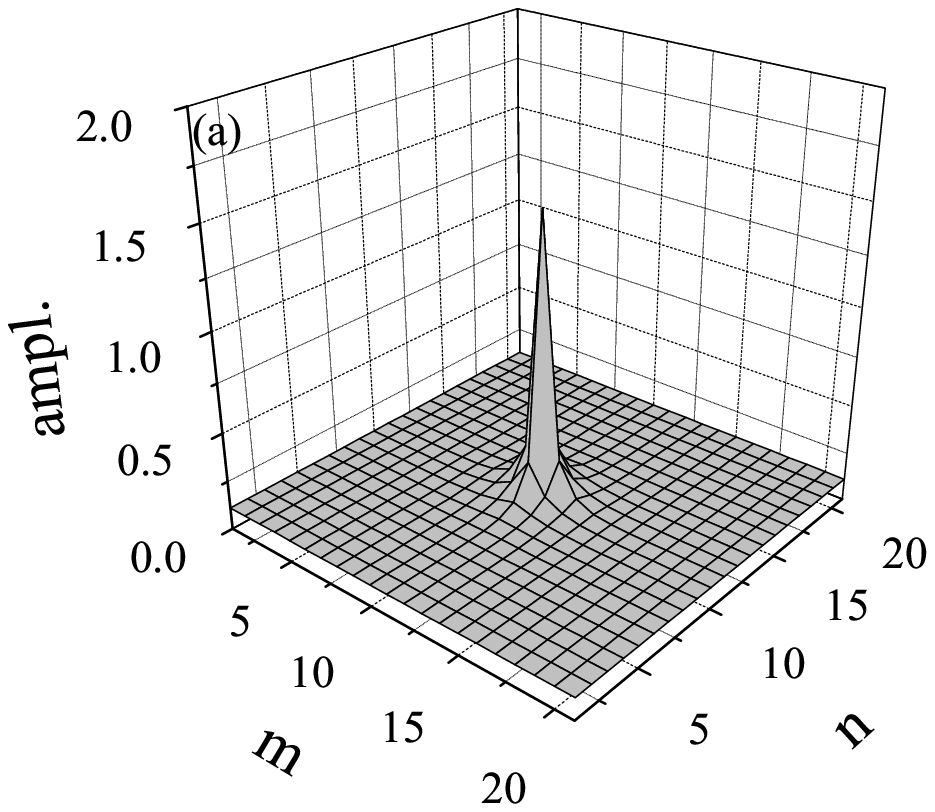}
\includegraphics [width=5cm]{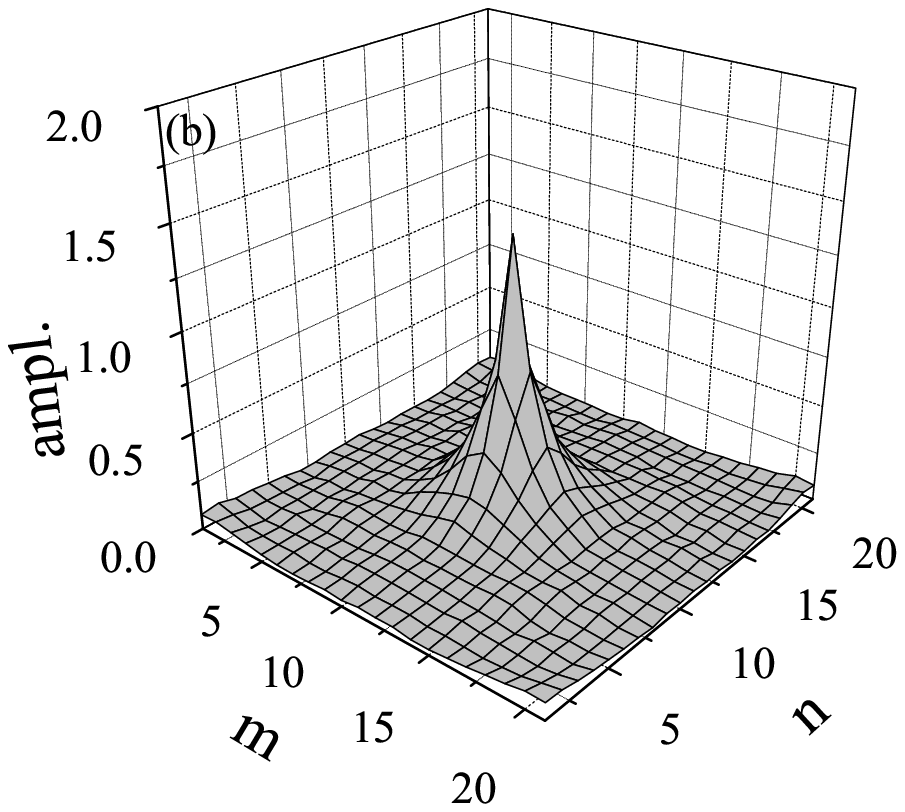}\includegraphics
[width=5cm]{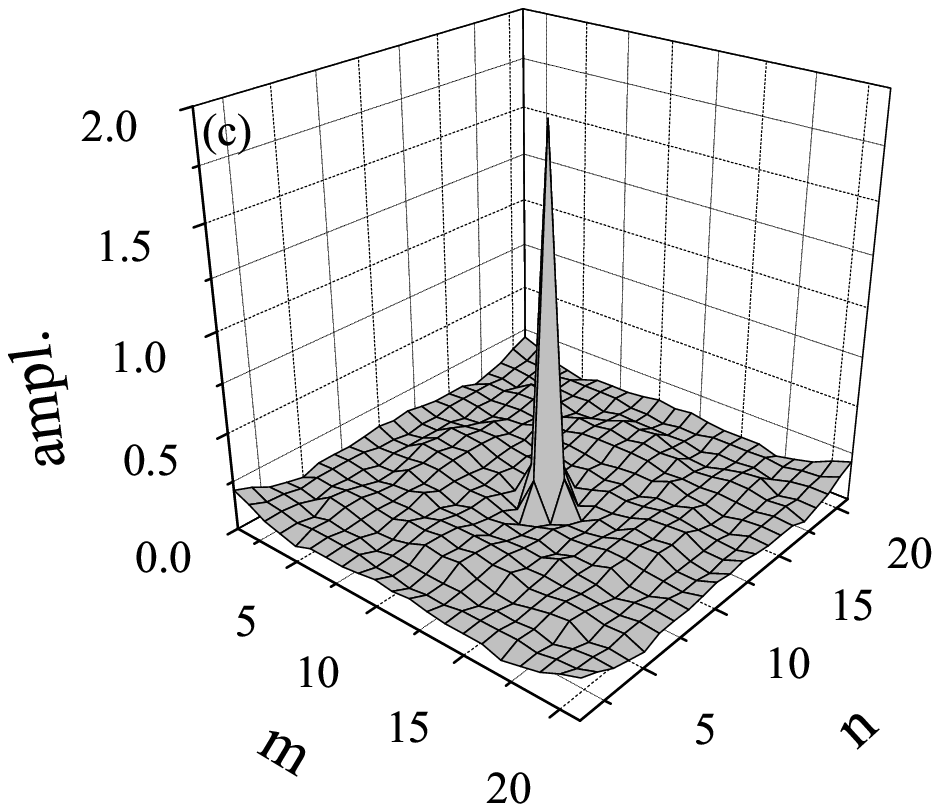}  \caption{The evolution of an unstable
perturbed soliton sitting on a finite background, for
$C=2,\protect\mu =1.1$, $\Gamma =1$, i.e., the combination of the
attractive contact and isotropic repulsive DD interactions.}
\label{fig11}
\end{figure}

The simulations demonstrate that the newly formed narrow breathers, built on
top of the oscillating background, are robust modes which survive any type
of perturbations. The intrinsic structure of these of the breathers depends
on the initial perturbations which initiated the formation of the breather
from an unstable soliton. Namely, for asymmetric or random perturbations,
the repulsion exerted by the DD interactions is more significant than in the
case of symmetric perturbations, making the emerging breather broader, with
a richer internal structure, see Fig. \ref{fig11}.

Finally, we have found that the fundamental solitons obtained in the
model combining attractive contact and \emph{anisotropic} DD
interactions with a finite strength (recall they correspond to the
in-plane orientation of the dipole moments) are unstable in their
entire existence region, except the on-site solitons in the region
of very small $C$ (the anti-continuum limit, as concerns the local
coupling). We could conclude that the anisotropy of the DD
interaction makes the solitons more sensitive to small perturbations
(i.e., more unstable), in comparison to the case of the isotropic DD
interactions.

\section{Fundamental solitons generated by the DD interactions in the case
of the repulsive local nonlinearity}

In the local model with the repulsive contact interactions ($\aleph =+1$),
unstaggered localized modes cannot exist. However, the presence of the DD
interactions, either attractive isotropic or anisotropic, opens a
possibility to create such modes, as recently shown in discrete 1D models
\cite{gpe,NPSE}, and was earlier predicted for quasi-2D continuous models in
Refs. \cite{Pedri05} (the competition of the repulsive contact and isotropic
attractive DD interactions) and \cite{Ami1} (the interplay of the repulsive
contact and anisotropic DD interactions).

We have found that the long-range DD interactions between lattice sites may
indeed create discrete 2D solitons in the case when the on-site nonlinearity
is repulsive. Numerical calculations show that, as in the situations
considered above, these discrete 2D solitons appear exactly in a part of the
parameter space where the CW states are modulationally unstable (this is
shown in detail in Ref. \cite{acta}). This region belongs to the
semi-infinite gap of the linear spectrum, which corresponds to $\mu <0$ and $%
\mu <1$, in the GPE and NPSE models, respectively. Again, three families of
fundamental unstaggered discrete solitons -- of the on-site, inter-site, and
hybrid types -- have been found in this situation as numerical solutions of
stationary equations (\ref{TdSNPSE}) and (\ref{TdSGPE}) for the DNPSE and
DGPE models, respectively.

It is relevant to emphasize that the solitons are also obtained in the limit
of the vanishing local interaction, i.e., the discrete solitons may be
supported solely by the DD interaction, which is relevant to the experiment
\cite{experim1}. In other words, the attractive nonlocal DD\ nonlinearity is
by itself sufficient for the formation of trapped localized modes in the 2D
lattices.

\subsection{Solitons generated by the attractive isotropic DD interaction}

In the case of the competition between the repulsive contact and
attractive isotropic DD interactions, all unstaggered fundamental
solitons have a bell-like shape, which is a consequence of the
nonlocality of the interaction which creates them. This feature is
also
similar to what was found in the 1D counterpart of the present models \cite%
{gpe}.

The $P\left( \mu \right) $ curves for the on-site, hybrid and inter-site
unstaggered solitons in the present case are much closer to each other than
it was in the case of the attractive contact interaction, see Fig. \ref%
{fig12}. On the other hand, numerical computations demonstrate a
violation of the spectral stability criterion for all types of the
unstaggered solitons in an almost entire existence region, except
for the case of a
small coupling constant, $C<0.1$, and certain values of $\mu $ and $\Gamma $%
. With the increase of the strength of the isotropic DD
attraction, the stability window shrinks with respect to the
values of $\mu $, as shown in Fig. \ref{fig13} for $C=0.1$.

Direct numerical simulations show that the unstable bell-shaped
localized structures evolve into the corresponding breathers, but,
on contrary to the situations considered above, \emph{without}
significant generation of an oscillating background, as shown in
Fig. \ref{fig14} (almost all the initial soliton's norm remains
localized, rather than being transferred to the background, unlike
the case of the attractive local interactions). The width of the
localized breather remains nearly the same as that of the original
unstable soliton.

\subsection{Solitons generated by the anisotropic DD interaction}

The anisotropic DD interaction can also provide for the formation
of unstaggered fundamental discrete solitons of the on-site,
hybrid and inter-site types in the case of the repulsive contact
interactions. Examples of the $P(\mu )$ characteristics for such
discrete solitons at fixed $\Gamma =5$ and two different values of
the coupling constant, $C=0.02$ (close to the anti-continuum
limit) and $C=2$ (close to the continuum limit), are shown in Fig.
\ref{fig15}. The respective $P(\mu )$ curves for solitons of the
on-site and hybrid types are very close to each other.

The linear-stability analysis indicates similar stability properties of the
on-site and hybrid solitons. They are unstable in almost the whole existence
region. As above, an exception is found close to the anti-continuum limit ($%
C<0.1$), where stable solitons exist. The respective stability
diagram is displayed in Fig. \ref{fig13} (b). In this figure, the
stable solitons of the on-site and hybrid types are found in the
white window. The stability window for the inter-site solitons
existing at $C<0.1$ is smaller than for their on-site and hybrid
counterparts.

We have also considered the limit of large values of coupling
constant $C$, i.e., the near-continuum limit, for which a narrow
stability region for solitons had been reported in Ref.
\cite{Ami1}, produced by the variational analysis and direct
simulations of the 3D continuous GPE. Taking into account the
geometry, stable quasi-2D solitary modes, characterized by
long-lived, slowly decaying oscillations were observed in certain
parameter regions. However, in the discrete 2D DGPE and DNPSE
models considered here (without the third dimension, that was
explicitly present in the analysis presented in Ref. \cite{Ami1}),
stable anisotropic solitons were not found in this part of the
existence region.

To summarize, in the model with the repulsive local interactions the
long-range attractive isotropic or anisotropic DD interactions give rise to
discrete solitons, but (marginally) stable ones can be found only close to
the anti-continuum limit. This conclusion may be considered as a consequence
of the long-range character of the DD interaction, which becomes more
significant as the system is getting more discrete \cite{Parker,Sanlew}.

\section{Conclusions}

In this paper we have introduced the 2D discrete model for the dipolar BEC
trapped in the deep optical lattice. We have analyzed the structure and
dynamics of fundamental discrete solitons in two versions of the model,
which are based, respectively, on the ordinary discrete form of the GPE
(Gross-Pitaevskii equation) with the cubic nonlinearity, or the 2D discrete
NPSE (nonpolynomial Schr\"{o}dinger equation), both including nonlocal DD
(dipole-dipole) interactions. Fundamental localized modes studied in this
work are counterparts of solitons existing in the semi-infinite gap, in
terms of the continuum limit of the discrete models. Only the situations
with \emph{competing} local and nonlocal interactions (attractive/repulsive,
or vice versa) appear to be interesting. We have shown close similarity in
the behavior of the fundamental unstaggered solitons in the discrete models
of both the GPE and NPSE types in these cases.

The presence of the repulsive isotropic DD interaction competing with the
local attraction extends the existence and stability regions for fundamental
on-site solitons. Due to the action of the nonlocal DD interactions, the
background plays an essential role in the formation of a new type of
unstaggered localized structures: near the edge of the semi-infinite gap of
the corresponding continuous system, the generation of SFBs (solitons on a
finite background) of the three types (on-site, inter-site and hybrid) was
revealed by the systematic numerical analysis. These solitons are unstable,
but they spontaneously transform themselves into robust breathers,
surrounded by a finite oscillating background. Actually, the presence of the
background is a feature specific to the dipolar model.

In the case of repulsive contact interactions, unstaggered fundamental
solitons may exist due to the DD attraction. In the case of the isotropic
attractive DD interaction, stable on-site, hybrid and inter-site solitons,
with close values of the norm, can be found in the strongly discrete limit.
In general, the fundamental solitons generated by the isotropic DD
interaction are broad bell-shaped modes. Being unstable, they evolve into
localized breathers with almost no loss of the norm.

\acknowledgments G.G., A.M., M. S. and Lj.H. acknowledge support from the
Ministry of Science, Serbia (Project 141034). The work of B.A.M. was
supported, in a part, by grant No. 149/2006 from the German-Israel
Foundation. This author appreciates hospitality of the Vin\v{c}a Institute
of Nuclear Sciences (Belgrade, Serbia).

\section{Figures}

\begin{figure}[tbp]
\center\includegraphics [width=7cm]{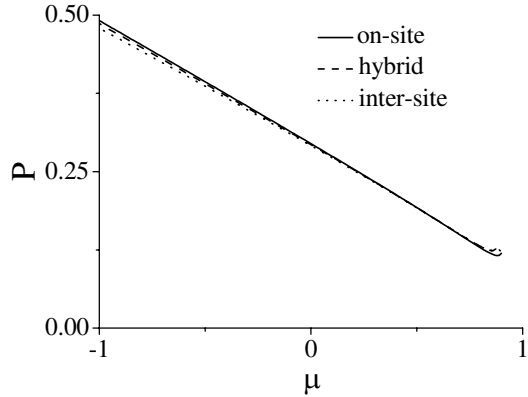} \caption{The
$P(\protect\mu )$ dependencies, for $C=0.8$ and $\Gamma =-10$, in
the model with the repulsive contact interaction and isotropic DD
attraction. For all the three types of the solitons -- on-site,
hybrid and inter-site -- the $P(\protect\mu )$ curves are very
close to each other.} \label{fig12}
\end{figure}

\begin{figure}[tbp]
\center\includegraphics [width=12.7cm]{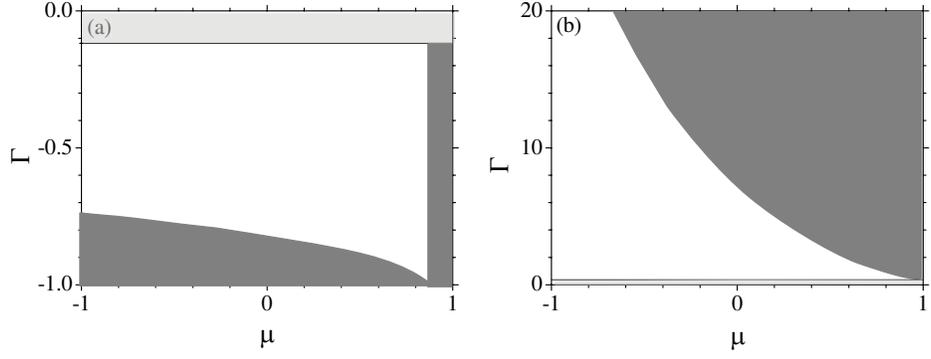} \caption{In
the case of the local repulsion, the white area is the stability
region for on-site unstaggered solitons with $C=0.1$, in the
presence of the isotropic DD attraction (a), or anisotropic DD
interaction (b). Inside the narrow light-gray areas, the
fundamental solitons do not exist.} \label{fig13}
\end{figure}

\begin{figure}[tbp]
\center\includegraphics [width=5cm]{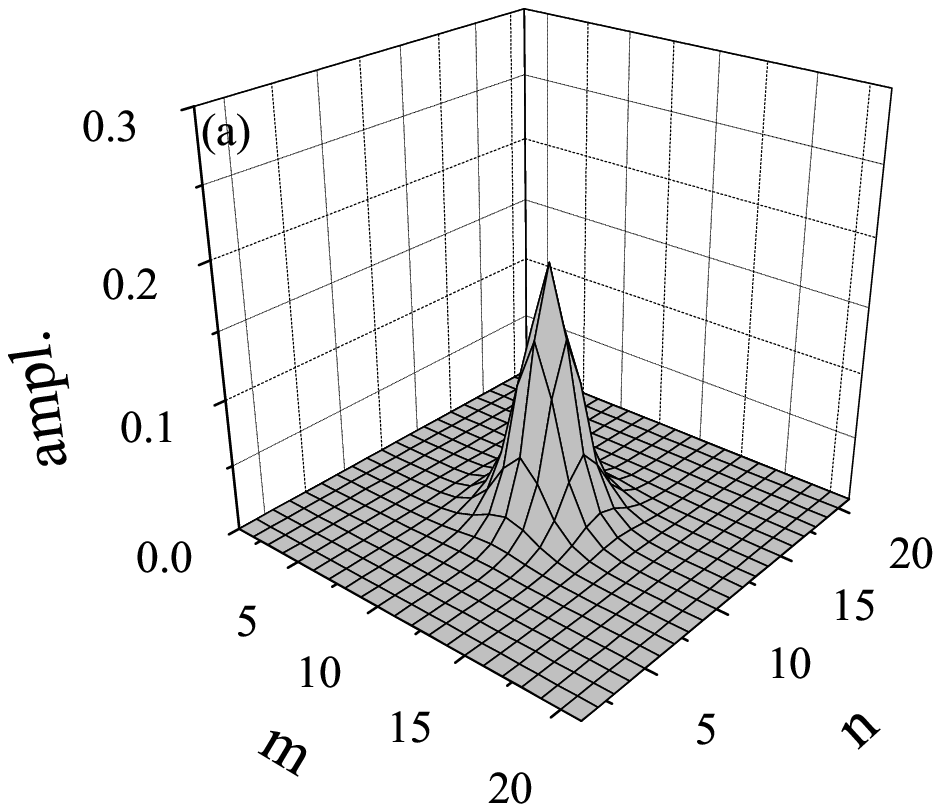}
\includegraphics [width=5cm]{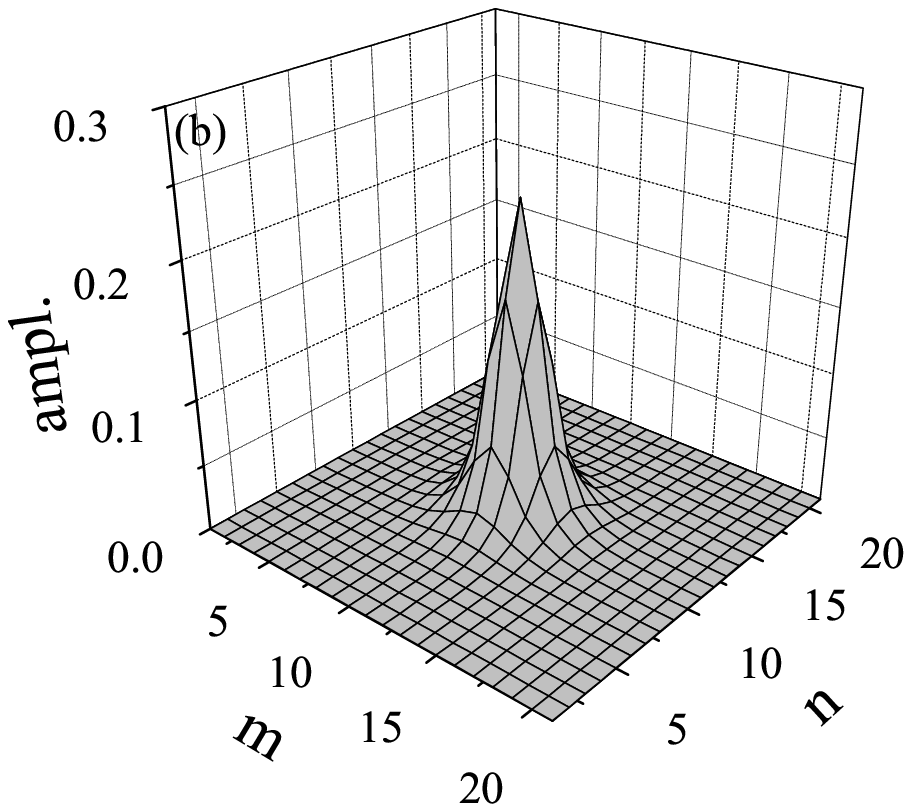}\includegraphics
[width=5cm]{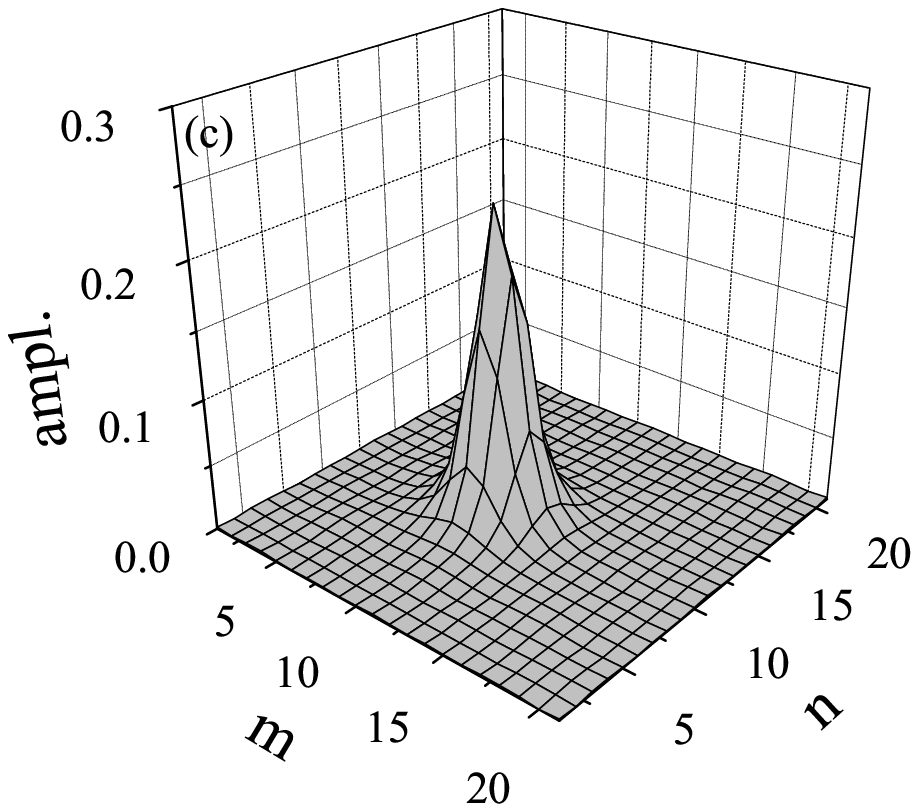} \caption{The evolution of a perturbed
bell-shaped on-site soliton with $C=0.8 $, $\protect\mu =0.5$,
$\Gamma =-10$, which corresponds to the combination of the on-site
repulsion and isotropic DD attraction. Plots present snapshots at
$t=1$ (a),  $t=10$ (b), and $t=60$ (c).} \label{fig14}
\end{figure}

\begin{figure}[tbp]
\center\includegraphics [width=12.7cm]{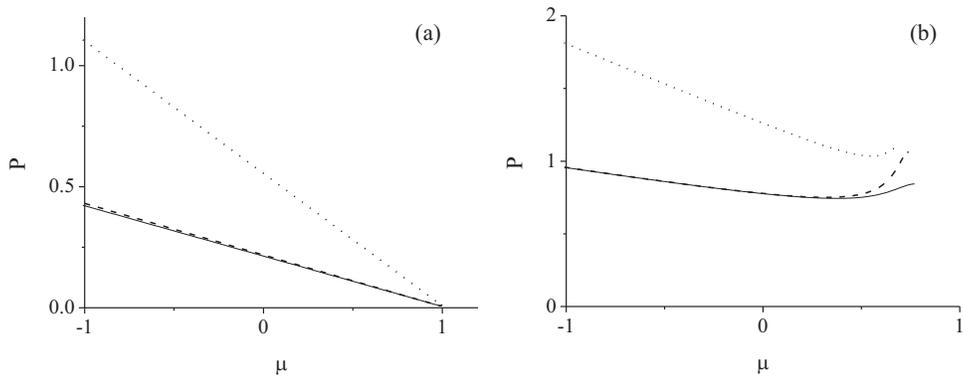} \caption{The
$P(\protect\mu )$ dependencies, for $C=0.02$ (a) and $C=2$ (b),
with fixed $\Gamma =5$, in the model with the repulsive contact
and anisotropic DD interactions. In both plots, the $P(\protect\mu
)$ curves for the three species of the discrete solitons are
shown: on-site (solid line), hybrid (dashed line) and inter-site
(dotted line).} \label{fig15}
\end{figure}

\end{document}